\documentstyle[prb,twocolumn,psfig,aps]{revtex}

\begin{document}

\draft
\twocolumn[\hsize\textwidth\columnwidth\hsize\csname @twocolumnfalse\endcsname

\title{Vortex-Loop Unbinding and Flux-Line Lattice Melting \\
in Superconductors}

\author{Anh Kiet Nguyen$^1$, A. Sudb{\o}$^1$, and R. E. Hetzel$^2$}
\address{
$^1$ ~ Department of Physics\\
Norwegian University of Science and Technology,
N-7034 Trondheim, Norway \\
$^2$ ~ Department of Theoretical Physics \\
Technical University of Dresden, D-01062 Dresden, Germany}
\maketitle

\begin{abstract}
We study the interplay  between a novel vortex-loop unbinding in finite
magnetic field at $T=T_V$ and flux-line lattice (FLL) melting at $T=T_M$
in type-II superconductors. The FLL melts due to nucleation of vortex
loops $\| \hat c$-axis, connected to flux lines. For moderate anisotropy,
phase coherence
$\| \hat c$ is lost at $T_V > T_M$ due to an $ab$-plane vortex-loop
unbinding with loops located close to thermal FLL fluctuations. For
large anisotropy, phase coherence $\| \hat c$ is lost at $T_V < T_M$
due to nucleation of $ab$-plane vortex-loops uncorrelated to flux lines.
\end{abstract}

\pacs{PACS numbers: 74.60.-w, 74.60.Ge}
\vskip2pc]
It is commonly assumed that the statistical mechanics of the flux-line
lattice (FLL) in type-II superconductors is mainly governed by fluctuations
of the flux lines around their positions in the ground state Abrikosov FLL,
at least at magnetic fields well below the upper critical field such that
vortex cores are well separated. Considerable progress in our understanding
of the FLL has been made over the last few years, under this assumption
\cite{Blatter:94}.

However, we show that for large mass-anisotropies the low-energy thermal
excitations in a superconductor in a low magnetic field are quite different.
There exists a set of important topological excitations, responsible for
destroying longitudinal superconducting phase coherence in a type-II
superconductor, which are qualitatively different from fluctuations of flux
lines. We also discuss in detail the fluctuations responsible for melting the
FLL with no pinning. These two distinct types of fluctuations may influence
each other in an essential way. This has
observable experimental consequences, as will be discussed. The induction
range considered is $0 < B/H_{c2} \ll 1$, which is known to be the relevant
limit for considering the phenomenon of FLL melting in extreme type-II
superconductors with uniaxial anisotropy \cite{Nelson:88,Houghton:89}.
Amplitude fluctuations of the superconducting order parameter are unimportant,
and the anisotropic London model is appropriate \cite{Sudbo:91a}.

The  model we consider is a discretized, uniaxial anisotropic superconductor
in the London limit \cite{Carneiro:93}, on a lattice with
$L_{\perp}^{2} \times L_z$ in units of the numerical lattice constant $d$.
The anisotropy is along the crystal $\hat c$-axis, described by the parameter
$\Gamma=\lambda_{c}/ \lambda_{a} = \sqrt{M_z/M}$, where $\lambda_{a}$ and
$\lambda_{c}$ are the magnetic penetration lengths along the crystal $ab$-plan
and the crystal $\hat c$-axis, respectively, and $M_z,M$ are normal-state
quasiparticle effective masses along the $z$-direction and $ab$-planes,
respectively. We take our coordinate system $(\hat x,\hat y,\hat z)$-axes
parallel to the crystal $\hat a, \hat b$ and $\hat c$-axes, respectively.
Periodic boundary conditions in all directions are used. The Hamiltonian is
given by \cite{Sudbo:91a,Carneiro:92}
\begin{equation}
\label{eq:Hamiltonian}
	H = 2 \pi^2 J_{0} \sum_{i,j} \sum_{\mu=x,y,z}
            G_{\mu}({\bf{r}}_{i}-{\bf{r}}_{j}) n_{\mu}({\bf{r}}_{i})
	    n_{\mu}({\bf{r}}_{j}),
\end{equation}
where $n_{\mu}({\bf{r}}_{i})$ is the integer vorticity through plaquette
$\mu$ at site ${\bf{r}}_{i}$. Note that this model allows for arbitrary
flux-line shapes, and also incorporates topological vortex loop excitations.
$J_{0} = \Phi_{0}^{2}d/16\pi^{3}\lambda_{a}^{2}$ and $J_{0}/\Gamma^{2}$ are
the energy scales for excitations involving $n_{z}({\bf{r}})$ and
$(n_{x}({\bf{r}}), n_{y}({\bf r }))$,
respectively, and $\Phi_{0}$ is the flux quantum.
G(${\bf{r}}$) is the lattice London interaction, with Fourier transform
\begin{eqnarray}
\label{eq:Gx-Gy}
	G_{x,y}({\bf{k}}) & = &
	                 \frac{\Gamma^{-2}}
			      {Q_x^2 + Q_y^2 + \Gamma^{-2}
			       (Q_z^2 + d^2 \lambda_a^{-2})}  \nonumber \\
	G_z({\bf{k}}) & = &
	       \frac{(Q^2 + d^2 \lambda_c^{-2})/(Q^2 + d^2 \lambda_a^{-2})}
			 {Q_x^2 + Q_y^2 + \Gamma^{-2}
			 (Q_z^2 + d^2 \lambda_a^{-2})},
\end{eqnarray}
where $Q_\mu=2 \sin(k_\mu d/2)$, $Q^2=\sum_\mu Q_\mu^2$.
In these simulations, we have chosen $\lambda_a=0.75 {\bar a}$,
where ${\bar a}$ is the average distance between flux lines.

To probe superconducting phase coherence, we consider the
helicity modulus $\Upsilon_\mu (q_\nu) \: (\mu \neq \nu)$
\cite{Teitel:95}. For a given perturbation in the
external vector potential $\delta A_{\mu}^{ext}(q_{\nu}) {\bf{\hat{\mu}}}$,
$\Upsilon_{\mu}(q_{\nu})$ is the linear response coefficient giving the
induced supercurrent ${\bf{j}}$
\begin{equation}
\label{eq:j}
	j_{\mu}(q_{\nu}) = - \Upsilon_{\mu}(q_{\nu},T)
			   \delta A_{\mu}^{ext}(q_{\nu}).
\end{equation}
We find that the helicity modulus in a  uniaxial anisotropic lattice
superconductor is given by
\begin{equation}
\label{eq:HelModXY}
\frac{\Upsilon_\mu (q_\nu,T)}
{\Upsilon_\mu (q_\nu,0)}
 = 1 - \frac{4 \pi^2 J_0 \lambda_a^2}
	  {VT}
     \frac{< n_\sigma (q_\nu) n_\sigma (-q_\nu) >}
          {1 + (1 + \delta_{\mu,z}(\Gamma^2-1))
           \lambda_a^2 Q^2},
\end{equation}
where $<..>$ denotes a thermal average, $(\mu, \nu, \sigma)$ are cyclic
permutation of $(x,y,z)$, and V is the volume of the lattice.

We investigate the melting of the FLL by considering the structure
factor for $n_z$ vortex elements
\cite{Carneiro:93}
\begin{equation}
\label{Structure Factor}
 S({\bf k})=\frac{<\mid \sum_i n_z({\bf{r}}_i)
                     \exp ~ [i {\bf k} \cdot {\bf r}_i]\mid^2>}
		    {N_z^2},
\end{equation}
where $N_z=\sum_i n_z({\bf{r}}_i)$. In the ground state the FLL has the well
know hexagonal form when ${\bf B} \| \hat c$, and  $S({\bf{K}},k_z=0)$
has $\delta$-function Bragg peaks at the reciprocal lattice points
${\bf{K}}$. When the FLL melts, the Bragg peaks are washed out. The
lowest $T$ where $S({\bf K})$ vanishes thus defines the melting temperature
$T_M$.

We employ the following Monte Carlo (MC) procedure: We start with
fixed average density f=B/$\Phi_{0}$ of straight vortex lines parallel to the
$\hat{z}$-axis. We update the system, heating from the ground state
by sweeping over the lattice in a systematic way. For each site we try to add
one elementary closed vortex loop, choosing one of six possible
orientations of the vortex loop  at random. These moves are accepted or
rejected according to the standard Metropolis algorithm. This MC procedure
provides a complete sampling of the phase space of the vortex variables
${\bf{n}}({\bf{r}})$, subject to two constraints: 1) ${\bf{\triangle}}
\cdot {\bf{B}}({\bf r}) =0$, where ${\bf{B}({\bf r})}$ is the local induction
and ${\bf{\triangle}}$ is the lattice gradient operator.  2) The average
induction is determined by ${\bf{B}} = (\Phi_0/V) \sum_i {\bf{n}}({\bf{r}}_i)
= f \Phi_0 \hat{z}$ is constant. Note that our MC procedure allows for
excitations of free vortex loops, i.e vortex loops not connected to any
existing vortex lines. Each data point is obtained after discarding the first
5000 steps to allow for equilibration, whereas the subsequent  10000-40000
steps are used to obtain averages, where each step refers to a sweep
through the entire $L_\perp^2 \times L_z$ lattice.

In these simulations, we use $f=1/48$, $L_{\perp}=24$ and $L_z=12$.
The reciprocal lattice vector of the vortex lattice ${\bf K}$ is taken
to be the lowest in $K_x$-direction consistent with the filling fraction
$f=1/48$ and a square numerical mesh, i.e. ${\bf K} =(4,0) 2 \pi/L_{\perp}$.
The $q$-value chosen in Eq. 4 is the lowest finite one on our numerical
lattice, i.e. ${\bf q}_x=(1,0) ~2 \pi/L_{\perp}$ and
${\bf q}_y=(0,1) ~ 2 \pi/L_{\perp}$.

\begin{figure}[htbp]
\centerline{\psfig{figure=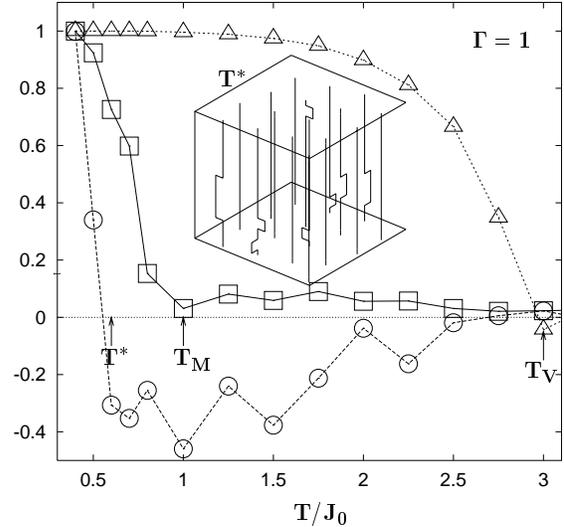,height=7.5cm,angle=270}}
\caption{$\Upsilon_z(q_x,T)/\Upsilon_z(q_x,0)$ ($\triangle$),
$\Upsilon_x(q_y,T)/\Upsilon_x(q_y,0)$ ($\bigcirc$), and $S({\bf K})$ ($\Box$)
as a function of temperature for $\Gamma=1$. Superconducting phase-coherence
along the direction of the flux lines is left intact in the flux-line liquid
$S({\bf K})=0$, indicating a melting transition into a disentangled vortex
liquid. Shown in the inset is a snapshot of the flux-line system at $T^*<T_M$.
All excitations are thermally nucleated flux-line defects. There are no free
vortex loops in the system, as explained in text. Note how $\Upsilon_x$
vanishes before $S({\bf K})$; the system is in a `floating solid phase'.}
\label{gamma1}
\end{figure}

\begin{figure}[htbp]
\centerline{\psfig{figure=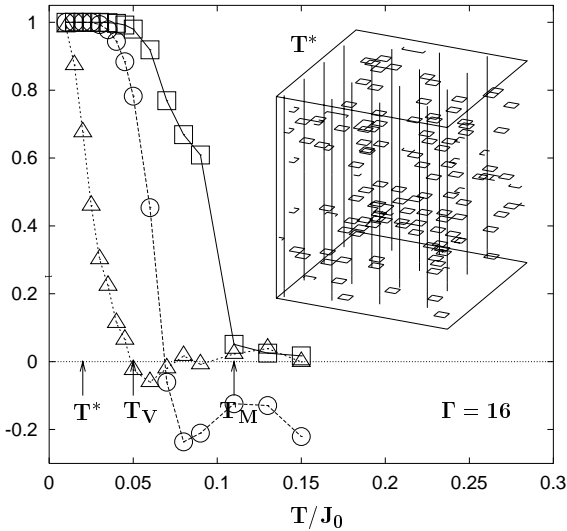,height=7.5cm,angle=270}}
\caption{$\Upsilon_z(q_x,T)/\Upsilon_z(q_x,0)$ ($\triangle$),
$\Upsilon_x(q_y,T)/\Upsilon_x(q_y,0)$ ($\bigcirc$), and $S({\bf K})$ ($\Box$)
as a function of temperature for $\Gamma=16$. Superconducting phase-coherence
along the direction of the flux lines is lost before the FLL melts. Shown in
the inset is a snapshot of the flux-line system at $T^* < T_M$. Note the
proliferation of $ab$-plane vortex loops uncorrelated to flux lines; the
FLL is left {\it intact}.}
\label{gamma2}
\end{figure}

$\Upsilon_{x,y}$ will always be zero in finite field when the FLL
is unpinned, since the FLL will move in response to an arbitrarily weak
applied current. In our lattice model, the numerical lattice introduces
artificial pinning which we must eliminate in order to be able to study the
true melting of the FLL. This is achieved for low enough values
of $f < f_c$, where  we have found $f_c \approx 1/32$ for the isotropic
systems and  decreasing very slightly with increasing mass-anisotropy. It
suffices that $\Upsilon_{x,y}$ vanishes at temperatures below
those of which $S({\bf K})$ vanishes. This indicates that the
FLL `floats' freely on the numerical lattice, having
thermally depinned from it, in the form of an intact FLL as demonstrated
by a finite $S({\bf K})$. This allows a study of the actual FLL-melting
without artificial pinning effects  \cite{Hattel:95}. As shown in both
Figs. \ref{gamma1} and \ref{gamma2}, this is clearly so in our
simulations. We have also checked that the $T$-dependence of $\Upsilon_z$,
practically is not influenced by $f$ if  $f < 1/16$.

The quantitity $\Upsilon_{z}(q)$ essentially measures phase-coherence along
the $\hat z$-axis. Since vortex loops need not be connected to flux lines,
$\Upsilon_{z}(q)$ may vanish without a corresponding vanishing of
$S({\bf K})$. {\it Thermal fluctuations may destroy superconducting phase
coherence while leaving the FLL intact}. This could possibly be interpreted
as the {\it finite-field} counterpart of the  well known `inverted
XY-transition' found in a $3D$ lattice superconductor model in {\it zero
magnetic field} \cite{Dasgupta:81}.

In Fig.~\ref{gamma1}, $S({\bf K})$, $\Upsilon_x$ and $\Upsilon_z$ are shown as
functions of the temperature for $\Gamma=1$. The helicity modulus $\Upsilon_z$
is seen to vanish at considerably higher temperatures than $S({\bf K})$.
Moreover, $\Upsilon_x$ vanishes at a temperature distinctly below $T_M$,
such that by the time we reach the FLL melting temperature $T=T_M$, we have
established a `floating solid phase' from which the FLL can melt with no
artificial pinning effects to the numerical lattice. The fact that
$\Upsilon_z$ stays finite through the melting transition where $S({\bf K})$
vanishes, indicates that the flux lines stay intact in the vortex liquid,
without large defects as evidenced by the still finite $\Upsilon_z$.
$S({\bf K})$ vanishes due to loss of shear stiffness between flux lines with
large effective tilt-moduli. As the temperature is increased further,
$\Upsilon_z$ eventually vanishes also, at $T=T_V$. For the isotropic case,
this is primarily due to fluctuations in the flux lines, or equivalently
directed vortex loops $\| \hat z$-axis {\it attached} to flux lines. This is
easily understood, since vortex loops $\| \hat z$-axis and vortex loops in the
$ab$-plane have the same self-energy when $\Gamma=1$, but vortex loops
$\| \hat z$-axis attached to flux lines gain energy by annihilating one
segment of the loop with one segment of a flux line. Moreover, when vortex
loops parallell to the $ab$-planes eventually are nucleated for
$\Gamma \alt 3$, they tend to be nucleated in close vicinity to flux-line
defects; they may interact attractively with the non-vertical (here
horizontal) components of these defects.

When $\Gamma > 5$, vortex loops in the $ab$-planes are easily nucleated
{\it anywhere} in the $ab$-plane, due to the large reduction of the
self-energy of flux lines per unit  length, by a factor $1/\Gamma^2$
\cite{Sudbo:91b}. The result is a dramatic enhancement of vortex-loop
unbinding and a resulting reduction of $T_V$ at which $\Upsilon_z(q)$
vanishes. This is shown in Fig. ~ \ref{gamma2}, where it is seen that, for
$\Gamma=16$, $\Upsilon_z$ vanishes at temperatures well below the temperature
$T_M$ where $S({\bf K})$ vanishes. $\Upsilon_z$ vanishes, not due to
fluctuations of the flux lines along the $\hat z$-axis, but due to nucleation
of vortex loops in the $ab$-planes located between flux lines. {\it We
conjecture that this transition  very likely may be  a finite-field
counterpart of the `inverted XY transition'  first discussed by Dasgupta
and Halperin of a lattice superconductor in zero-field} \cite{Dasgupta:81}.
Flux-line defects are nucleated only at much higher temperatures, and the
FLL is left completely intact at the vortex-loop unbinding transition, as
shown by the value of $S({\bf K})$ when $\Upsilon_z$ vanishes. This is also
clearly illustrated in the inset of Fig. \ref{gamma2}.

Vortex-loops $ \| \hat z$-axis also become easier to nucleate when $\Gamma$
increases, but only by half the amount of the $ab$-plane loops, since an
elementary loop $\| \hat z$-axis has two, and not four, flux-line segments
parallell to the $ab$-plane. Moreover, it is always energetically favourable
to attach them to already existing flux lines, by the same argument as in the
isotropic case. Hence, flux-line defects are also easier to nucleate in the
anisotropic case, as one would have guessed from the reduction of the nonlocal
tilt-modulus of the FLL when $\Gamma$ increases \cite{Houghton:89}. Practically
no vortex loops $\| \hat z$-axis are nucleated away from existing flux lines
for $T \leq T_M$.

For sufficiently large $\Gamma$, $\Upsilon_z$ vanishes at temperatures well
below those at which $S({\bf K})$ vanishes, i.e. $T_V \ll T_M$. Nonetheless,
the melting of the FLL is not necessarily $2D$.  $T_M(\Gamma)$ continues to
drop for increasing values of $\Gamma$ which are much larger than the
value of $\Gamma$ where $T_V=T_M$. Thus, even if $T_V \ll T_M$, $T_M$
still depends on the coupling between planes. As long as this is the case,
the FLL melting transition is therefore $3D$ in character. This is shown in
Fig. 3, where both $T_V$ and $T_M$ are plotted versus $\Gamma$. $T_V=T_M$
for $\Gamma \approx 5$. However, $T_M$ only saturates to a finite value (which
implies $2D$ melting \cite{Doniach:79}) at a much larger value,
$\Gamma \approx 16$. We have found an intermediate range of $\Gamma$-values
where the entropy jump $\Delta S$ associated with the FLL melting increases
with $\Gamma$ \cite{Nguyen:96}.

\begin{figure}[htbp]
\centerline{\psfig{figure=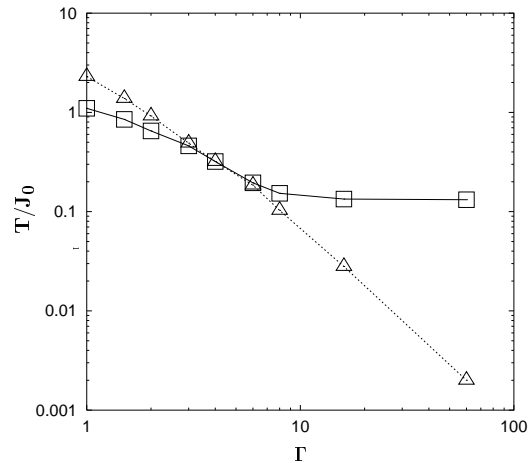,height=6.5cm,angle=270}}
\caption{Melting temperature $T_M$ ($\Box$) and vortex-loop unbinding
temperature
$T_V$ ($\triangle$)
as a function of $\Gamma$. Note saturation at $\Gamma > 16$, showing
that the melting eventually becomes $2D$, but only for considerably larger
values of $\Gamma$ than those values where $T_M=T_V$.}
\label{tmbmz}
\end{figure}

Previously, an entropy jump of $\Delta S = 0.3 k_B$ per vortex per layer
was found theoretically \cite{Hetzel:92}, well in agreement with a
number of experiments \cite{Pastoriza:94,Zeldov:95}. We now propose a
picture for the variation $\Delta S$ with $\Gamma$ and induction $B$,
when the perfect Abrikosov FLL has been gauged away by a singular
gauge-transformation \cite{Tesanovic:95}. {\it Hence, only thermally
nucleated vortex-loops and thermally nucleated flux-line defects are
relevant degrees of freedom, precisely as in our Monte Carlo simulations}.

The character of the vortex-loop unbinding and FLL melting transitions is
governed by the effective interactions between vortex-loops in the $ab$-planes
and the interaction between flux-line defects, respectively. {\it The
effective interaction between vortex loops is screened by thermally nucleated
flux-line defects, but is left unscreened by the flux lines in the rigid
Abrikosov vortex lattice} \cite{Dasgupta:81,Tesanovic:95}.  Conversely,
flux-line defects are screened by vortex loops. Note that this is different
from the electromagnetic screening of the interaction between field-induced
vortices, even in the perfect Abrikosov FLL. Such screening is present at all
temperatures.

\underline{Vortex-loop unbinding transition}:
An unscreened interaction between vortex loops \cite{Dasgupta:81,Tesanovic:95}
makes the vortex-loop unbinding transition  continuous, while screening will
tend to drive it first order. Increasing $\Gamma$ will enhance thermal
nucleation of flux-line defects, as argued above, enhancing screening of the
vortex-loop interaction, driving the vortex-loop unbinding more first order.
If one increases $B$, an increase in the density of flux-line defects will
also enhance screening of the effective interaction between vortex loops,
enhancing the first order character of the transition. However, it will also
suppress fluctuations in individual flux lines, by increasing the flux-line
tilt modulus $c_{44}$. At low $B$ $c_{44} \sim B^2$, while the areal density
of flux-lines $n \sim B$. Therefore, the dominant effect is the increase of
the areal density of flux-line defects. Screening of the vortex-loop
interaction is increased, driving the vortex-loop unbinding more
discontinuous. There will be a crossover field $B^*$ above which the increase
in $c_{44}$ dominates, which will then reduce the number of flux-line defects.
Above this field, the vortex-loop unbinding transition will become less first
order.

\underline {FLL melting transition}:
An unscreened interaction between flux-line defects will make the FLL
melting transition continuous. Screening of this interaction will tend to
drive it first order. Increasing $\Gamma$ will increase the nucleation of
vortex loops in the $ab$-plane, enhancing the screening of  the interaction
between flux-line defects, driving the melting of the FLL more first order.
On the other hand, if the magnetic field increases, it is clear that the
interaction between thermally nucleated flux-line defects will be less
screened, driving the transition more continuous \cite{Tesanovic:96}. The FLL
melting transition thus becomes less first order in the anisotropic case as
$B$ increases, due to the presence of already thermally nucleated $ab$-plane
vortex loops. Therefore the entropy jump associated with the transition is
reduced as $B$ increases.

Recently, an interesting decrease in the entropy jump $\Delta S$ with
increasing $B$ was in fact observed by Zeldov {\it et al.} \cite{Zeldov:95}.
Note that a consideration of fluctuations of the flux lines  only, will give
$\Delta S \sim B^{1/2}$, an opposite trend \cite{Zeldov:95}. The above
discussion shows that inclusion of vortex-loops into the consideration of
the FLL melting in a natural way explains the data of Zeldov {\it et al.}.
The vortex-loop unbinding is crucial for the statistical mechanics of a
type-II superconductor in  magnetic fields $B \in (0-2T)$.

We have discussed the existence of, and interplay between, two distinct
phase-transitions in type-II superconductors. One is the melting of the FLL,
which looses its meaning as the induction goes to zero. The other is a novel
vortex-loop unbinding transition in finite magnetic field, with a zero-field
counterpart. We emphasize two novel aspects of this work. The vortex-loop
unbinding transition has been observed i) in a finite magnetic field and
ii) both in the flux-line lattice and flux-line liquid cases, depending on
our choice of $\Gamma$.

Support from the Research Council of Norway under Grants No. 110566/410
and 110569/410, is gratefully acknowledged. We thank K. Fossheim,
F. Kusmartsev, Z. Te{\v s}anovi{\'c}, and J. M. Wheatley for discussions,
and F. Mo for use of computational facilities.


\begin{references}
\bibitem{Nguyen:96} Calculations of the entropy discontinuity both at the
vortex-loop unbinding transition and the FLL melting transition, as a
function of $\Gamma$, are in progress. In this paper $L_z=12$. Simulations
for $L_z=6,12,18,24$ reveal  that for $\Gamma=1$, the transitions become
sharper, while for $\Gamma=16$ the transitions are unaffected by increasing
$L_z$.
\end{references}
\end{document}